\documentclass[12pt]{article}
\usepackage{fps}
\newcommand{\rf}[1]{(\ref{#1})}
\begin{document}

\begin{flushright}
HU-EP-01/58 \\
DPNU-01-34
\end{flushright}

\vspace*{2mm}

\begin{center}

{\LARGE\bf The Area Law in Matrix Models \\
\vspace*{6mm} 
for Large $N$ QCD Strings}

\vspace*{10mm}

K.N. Anagnostopoulos $^{\rm a}$\footnote
{e-mail address : konstant@physics.uoc.gr},
W. Bietenholz $^{\rm b}$\footnote
{e-mail address : bietenho@physik.hu-berlin.de} and 
J. Nishimura $^{\rm cd}$\footnote
{e-mail address : nisimura@eken.phys.nagoya-u.ac.jp}

\vspace*{6mm}

$^{\rm a}$ Dept. of Physics, University of Crete \\
P.O. Box 2208, GR-71003 Heraklion, Greece \\

\vspace*{4mm}

$^{\rm b}$ Institut f\"{u}r Physik, 
Humboldt Universit\"{a}t zu Berlin \\
Invalidenstr. 110, D-10115 Berlin, Germany \\

\vspace*{4mm}

$^{\rm c}$ The Niels Bohr Institute, Blegdamsvej 17 \\
DK-2100 K\o benhavn \O , Denmark \\

\vspace*{4mm}

$^{\rm d}$ Dept.\ of Physics, Nagoya University \\
Nagoya 464-8602, Japan

\end{center}

\vspace*{6mm}

{\it We study the question whether matrix models obtained in
the zero volume limit of 4d Yang-Mills theories can describe
large $N$ QCD strings. 
The matrix model we use is a variant of the Eguchi-Kawai model in
terms of Hermitian matrices, but without any twists or quenching. This
model was originally proposed as a toy model of the IIB matrix
model. In contrast to common expectations, we do observe the area law
for Wilson loops in a significant range of scale of the loop area.
Numerical simulations show that this range is stable as $N$
increases up to 768, which strongly suggests that it persists in the
large $N$ limit. Hence the equivalence to QCD strings may hold for
length scales inside a finite regime.}

\newpage

\section{Eguchi-Kawai equivalence}

In quantum field theories and systems in statistical mechanics,
the number of internal degrees of freedom often enters as a
free parameter, which we denote by $N$.
In many cases, the dynamics of the system simplifies
significantly as $N$ becomes large. A general discussion is given 
for instance in Ref.\ \cite{Das}.

This effect motivated in particular the model suggested
by Eguchi and Kawai a long time ago \cite{EK82}.
Their point of departure was standard $U(N)$ lattice gauge
theory. Based on the factorization of 
correlation functions
at $N \to \infty$, they suggested that the model should be equivalent
to its dimensional reduction to one point. Then all the link
variables are replaced by $U_{x,\mu} \to U_{\mu}$, and the
plaquette action reduces to
\begin{equation}
\label{eq1}
S_{EK} = -N\beta \sum_{\mu \neq \nu =1}^{d}
{\rm Tr} (U_{\mu} U_{\nu} U_{\mu}^{\dagger} U_{\nu}^{\dagger}) \ .
\end{equation}
As their argument for the {\em Eguchi-Kawai equivalence} to ordinary
gauge theory, the authors showed that the Schwinger-Dyson equations
remain unaltered, which also implies the invariance of the Wilson
loops. However, their derivation implicitly assumed that the
$[U(1)]^{d}$ symmetry of the phases is not spontaneously broken. This
is correct at strong coupling in dimension $>2$ and for all couplings
in two dimensions. On the other hand, it turned out that at weak
coupling in dimension $>2$, spontaneous symmetry breaking does set
in. Numerical simulations revealed that the $N$ eigenvalues of
$U_{\mu}$ tend to be almost equal \cite{SCRI}. Hence this attempt to
prove general equivalence to standard lattice gauge theory failed, and
therefore the equivalence to the continuum gauge theory is uncertain
as well.

In order to avoid this problem, soon afterwards modified versions of
the Eguchi-Kawai model were introduced, in particular the ``quenched
Eguchi-Kawai model'' \cite{SCRI} and the ``twisted Eguchi-Kawai
model'' \cite{GAO}, where the phase symmetries are not spontaneously
broken. Ref.\ \cite{SCRI} also contains numerical results on the
quenched Eguchi-Kawai model, and Refs.\ \cite{Parisi,GrKi} present a
perturbative consideration.  For numerical studies and generalizations
of the twisted Eguchi-Kawai model, see for instance Ref.\ \cite{FH}.\\

More recently, Ishibashi, Kawai, Kitazawa and Tsuchiya worked out
a large $N$ reduced matrix model of 10d super Yang-Mills theory
\cite{IKKT} (for a review, see Ref.\ \cite{Aoki:1998bq}).
The authors presented some evidence for its relation to type IIB
superstring theory and conjectured that it could provide a
constructive (non-perturbative) definition of string theory.  In
particular, they showed that the matrix model action reduces to the
Green-Schwarz-Schild action in a certain semiclassical limit
\footnote{A more in depth study of this correspondence has been
reported in Refs.\ \cite{Anagnostopoulos:2000mn}.}.
Moreover, their {\em IIB matrix model} is shown to
describe D-branes and their interactions correctly,
which was further elaborated in Refs.\ \cite{Dbranes}. 

Supersymmetry seems to protect the $[U(1)]^{d}$ symmetry \cite{IKKT},
but it turns out to be marginally broken
\cite{AIKKT}.
It is completely broken in the simplified {\em bosonic IIB 
matrix model}, which is obtained by omitting
the fermions.
Nevertheless, the bosonic model has been recognized as a well-defined
model \cite{KS}, and it has attracted considerable interest
in the literature.  Both, the bosonic and the supersymmetric model can
be formally obtained from the zero-volume limit of large $N$ gauge
theories in the continuum, and hence they can be viewed as variants
of the original Eguchi-Kawai model.  The finiteness of the partition
function becomes a non-trivial issue, since the matrices to be
integrated are Hermitian.  The condition for the finiteness ($N \ge
2$, $d=4,6,10$ for the supersymmetric case, and
$N>\frac{3d-4}{2(d-2)}$ for the bosonic case) conjectured in Ref.\
\cite{KS,KNS} has been proved recently \cite{welldef}.  However, the
validity of Eguchi-Kawai equivalence is still an open question.

In particular, the 4d case of this Hermitian matrix model is
accessible for numerical studies.  This is true even for its
supersymmetric form, because the fermionic determinant is real
positive in $d=4$ \cite{AABHN}, as suspected earlier \cite{KNS}.
\footnote{The bosonic model has also been simulated in higher
dimensions. In particular, Ref.\ \cite{HNT} presents results up
to $d=20$, $N=32$, which are compared to a $1/d$ expansion,
and Ref.\ \cite{HoEg} arrives at $d=10$, $N=128$.
Also the SUSY model without fermionic phase factor was simulated
in $d=6$ and $d=10$ \cite{nophase}.}
Large $N$ factorization is known to hold in the 4d IIB matrix
model, both, for the bosonic \cite{HNT}
and for the supersymmetric case \cite{AABHN}.
As a stronger criterion for Eguchi-Kawai equivalence to ordinary
gauge theory, one may investigate the validity of the area law for
the Wilson one-point function.
Simulations for $N$ up to 48 show a finite range of the scale,
where the area law seems to hold \cite{AABHN}.
\footnote{A similar behavior was also observed in the 10d bosonic
model \cite{HoEg}.} 
For the SUSY case, larger $N$
can hardly be simulated on present computers, but in the bosonic
case it is possible to go far beyond. 
\footnote{The computational effort grows like $N^{5}$ in the
SUSY case, but only like $N^{3}$ in the bosonic model \cite{AABHN}.}
The goal of this work
is to investigate the fate of the finite area law window 
as $N$ becomes really large. We increase $N$ up to 768, and observe
that the scaling window does indeed stabilize;
this provides strong evidence that it
does neither shrink to zero nor extend to a larger regime
in the large $N$ limit. We also confirm for $N$ up to 768 that the
correlation functions relevant for QCD and string theory remain finite
in the large $N$ limit if one keeps the product $g^2N$ fixed
\cite{AABHN}. 

In Section 2 we describe the model we are studying, and
the properties which are relevant in this context.
In Section 3 we recall the notion of Wilson loops in the framework
of matrix models and their scaling behavior. Then we present
our results for the area law at $N=48$, 128, 256, 512 and 768
for the bosonic 4d IIB matrix model.
In Section 4 we discuss the conclusions from these results, 
which were mentioned before in Refs.\ \cite{proc}.


\section{The model}


In 1997, Ishibashi, Kawai, Kitazawa and Tsuchiya suggested a 
model of supersymmetric matrices, which is a candidate for a
non-perturbative definition of IIB superstring theory \cite{IKKT}.
Here we are interested in the bosonic variant of this model, which is
obtained if one simply drops the fermions by hand.  The resulting {\em
bosonic IIB matrix model} is given by the partition function
\begin{eqnarray}
Z &=& \int dA \ \exp (-S[A]) \ , \\
S[A] &=& - \frac{1}{4g^{2}} {\rm Tr} \, \Big( [A_{\mu}, A_{\nu}]^{2}\Big) 
\ , \nonumber
\end{eqnarray}
where $A_{\mu}$ are traceless Hermitian $N \times N$ matrices, and in
the $d$ dimensional version of this model, $\mu$ runs from 1 to
$d$. The original model was formulated in $d=10$, but here we consider
$d=4$.  The constraint of tracelessness is incorporated in the
measure,
\begin{equation}
dA = \prod_{\mu = 1}^{d} \left[ \prod_{i>j} \{ d {\rm Re}
(A_{\mu})_{ij} \ d {\rm Im} (A_{\mu})_{ij} \} 
\prod_{i=1}^{N}
d (A_{\mu})_{ii} \ \delta\Big( \sum_{i=1}^{N} 
(A_{\mu})_{ii} \Big) \right] \ .
\end{equation}
Note that this model is well-defined without any cutoff for 
$N>\frac{3d-4}{2(d-2)}$ 
\cite{KS,welldef}.
Hence its only parameter $g$ is a pure scaling parameter, rather than
a coupling constant.  It can simply be absorbed by rescaling the
variables as
\begin{equation}
A_{\mu} = \sqrt{g} \, X_{\mu} \ ,
\end{equation}
where the matrices $X_{\mu}$ are dimensionless.

This model is invariant under Euclidean rotations of the $A_{\mu}$,
and it has the $SU(N)$ symmetry
\begin{equation}
A_{\mu} \rightarrow V A_{\mu} V^{\dagger} \ , \qquad
V \in SU(N) \ .
\end{equation}


This 4d bosonic IIB matrix model is closely related to the original
Eguchi-Kawai model in the weak coupling limit
(for details, see e.g.\ Ref.\ \cite{HNT}).
Due to the $[U(1)]^d$ spontaneous symmetry breaking the eigenvalues 
of the model \rf{eq1} collapse to a point and we obtain
\begin{equation}
\label{eka}
S_{EK}\approx -\frac{1}{2}N\beta\sum_{\mu\neq\nu=1}^{d} {\rm
Tr}\Big( [A_\mu,A_\nu]^2\Big) +{\cal O}(A^6)\, ,
\end{equation}
where $U_\mu = \mbox{e} ^{i A_\mu}$.
Thus we can identify $\beta=1/(2g^2N)$ and the two
models agree in the limit $\beta\to\infty$. 

Moreover, the large $N$ limit of the bosonic model
is expected to be equi-valent to the 't Hooft limit of the 
corresponding Yang-Mills theory with a certain 
't Hooft coupling constant $\lambda_{\rm YM}$ and 
a momentum cutoff $\Lambda$ \cite{HNT}.
In this correspondence, however, the 't Hooft coupling constant
is related to the momentum cutoff canonically, and therefore
one cannot tune $\lambda_{\rm YM}$ appropriately
as one sends $\Lambda \to \infty$.

In the next Section we are going to take a new look at the possibility
of the Eguchi-Kawai equivalence of the bosonic model to ordinary gauge
theory at large $N$.  After all, the presence of spontaneous symmetry
breaking does not rule out that Eguchi-Kawai equivalence could be
realized for some range of length scales. As a powerful criterion to
test this possibility, we study 
how far the Wilson loops follow an area law.

\section{The area law}

In analogy to ordinary gauge theory, Polyakov lines
and Wilson loops are introduced as
\begin{eqnarray}
P(k) &=& \frac{1}{N} {\rm Tr} [ \exp (ikX_{1}) ] \ , \quad {\rm and}\\
W(k) &=& \frac{1}{N} {\rm Tr} [ \exp (ikX_{1}) \exp (ikX_{2})
\exp (-ikX_{1})\exp (-ikX_{2}) ] \ ,
\end{eqnarray}
respectively. For convenience we insert particular components
of the dimensionless matrices $X_{\mu}$, but their choice
is irrelevant. 
In the interpretation of large $N$ reduced matrix models as string
theory, Wilson loops play the r\^{o}le of string creation or annihilation
operators \cite{FKKT}. 
The parameter $k$ represents the momentum distribution carried by
each section of the string.
Its physical (dimensionful) counterpart is given by
$k_{\rm phys} = k / \sqrt{g}$.
Correlation functions of these operators
have a direct physical meaning, and hence the existence of a 
non-trivial large-$N$ limit is absolutely crucial. 
It has been observed to exist if one
tunes $g \propto 1 / \sqrt{N}$, in the bosonic and in the
supersymmetric case \cite{AABHN}. In what follows, we set
$g =\sqrt{48/N}$ following the convention in Ref.\ \cite{AABHN}.
In the Eguchi-Kawai equivalence, 
$\langle W(k)\rangle$ corresponds to the Wilson loop in the
large $N$ gauge theory, and the parameter $k_{\rm phys}$ represents 
the linear extent of the Wilson loop in a physical scale.

\begin{figure}[hbt]
\def\fpsangle{270}
\epsfxsize=100mm
\fpsbox{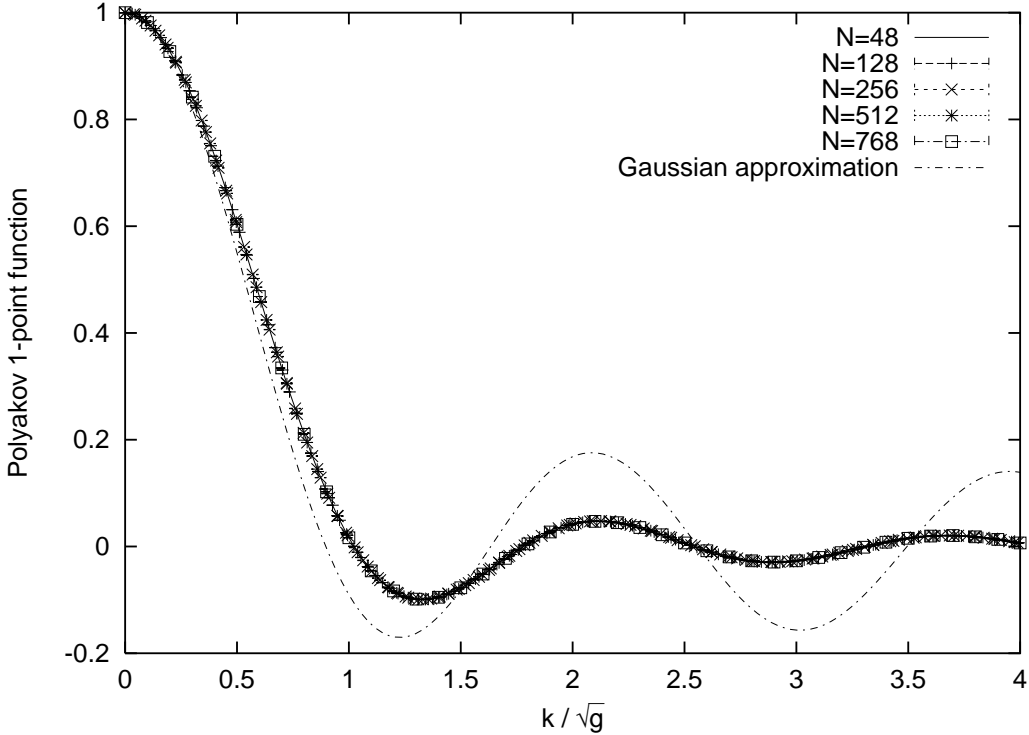}
\caption{\it{The Polyakov line $\langle P \rangle$ plotted 
against the dimensionful momentum $k_{\rm phys}=
k / \sqrt{g}$. Numerical results for $N=48 \dots 768$
confirm the large $N$ scaling, as well as qualitative agreement with
the Gaussian prediction.
}}
\label{poly-fig}
\end{figure}

In Fig.\ \ref{poly-fig} we illustrate the large $N$ scaling further by
showing the Polyakov line $\langle P \rangle$ as a function of $k_{\rm
phys}$, with $N=48$, 128, 256, 512 and 768. Scaling is
confirmed in an impressive way. 
We also observe that the $k_{\rm phys}$-dependence of the 
scaling function is in qualitative agreement
with an analytical prediction by Oda
and Sugino, which is based on a Gaussian approximation (to the
second order) in the large $N$ limit \cite{Sugino}.
The observation that $\langle P \rangle$ does not vanish at finite
values of $k_{\rm phys}$ confirms that the $[U(1)]^{4}$ symmetry is
broken, as we mentioned in Section 1.

\begin{figure}[hbt]
\def\fpsangle{270}
\epsfxsize=100mm
\fpsbox{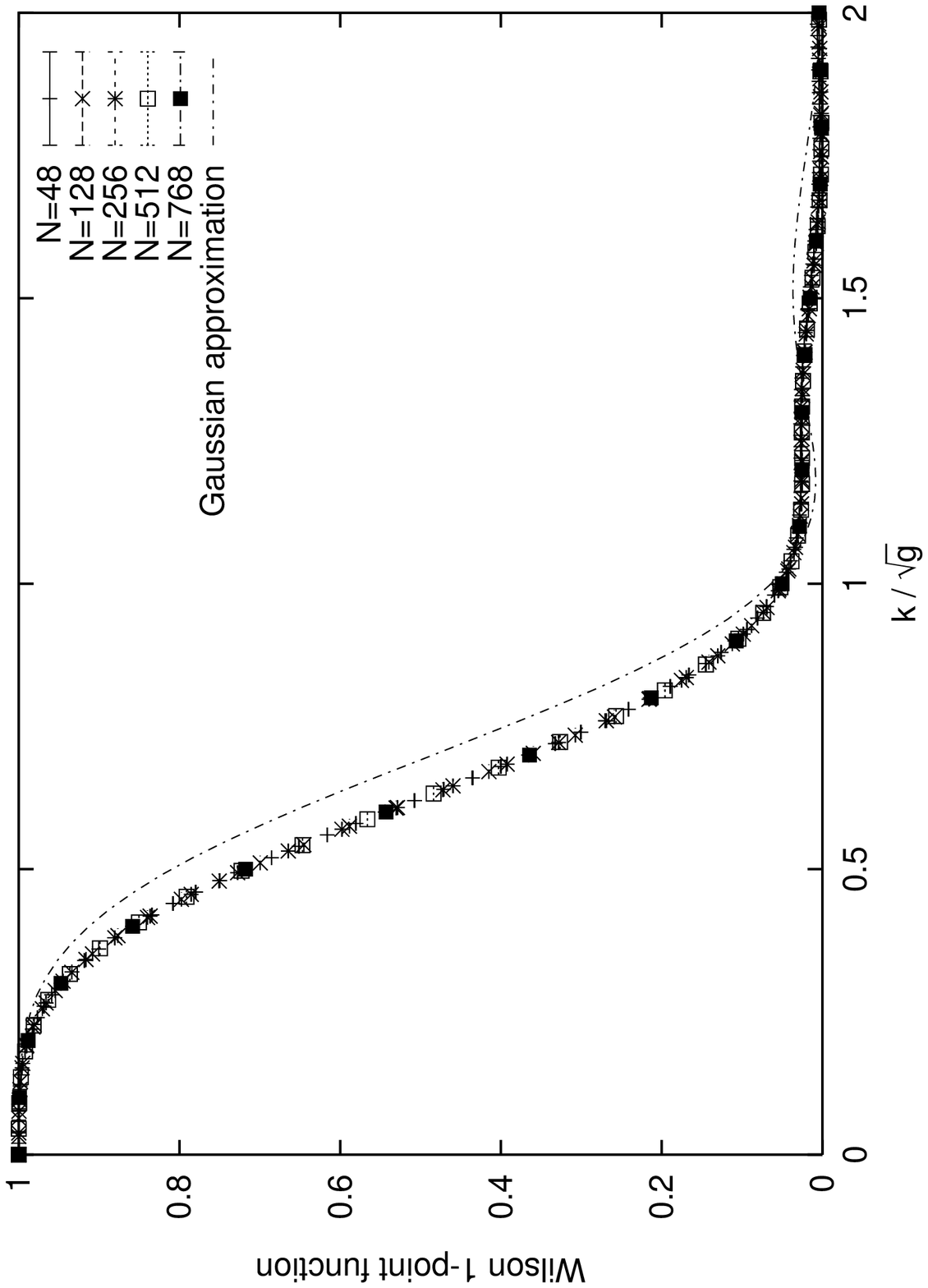}
\caption{\it{The Wilson loop $\langle W \rangle$ plotted against the
dimensionful momentum $k_{\rm phys}= k / \sqrt{g}$. Numerical results
for $N=48 \dots 768$ confirm the large $N$ scaling. Again the Gaussian
approximation is shown for comparison.}}
\label{wil-fig}
\end{figure}

Fig.\ \ref{wil-fig} is the corresponding plot for the Wilson loop
$\langle W \rangle$,
also as a function of $k_{\rm phys}$. We see that large $N$ scaling 
is confirmed again, hence we can take a rather safe extrapolation
to $N \to \infty$. Again we show the Gaussian approximation 
(to the first order) \cite{Sugino}
for comparison. 

\begin{figure}[hbt]
\def\fpsangle{270}
\epsfxsize=100mm
\fpsbox{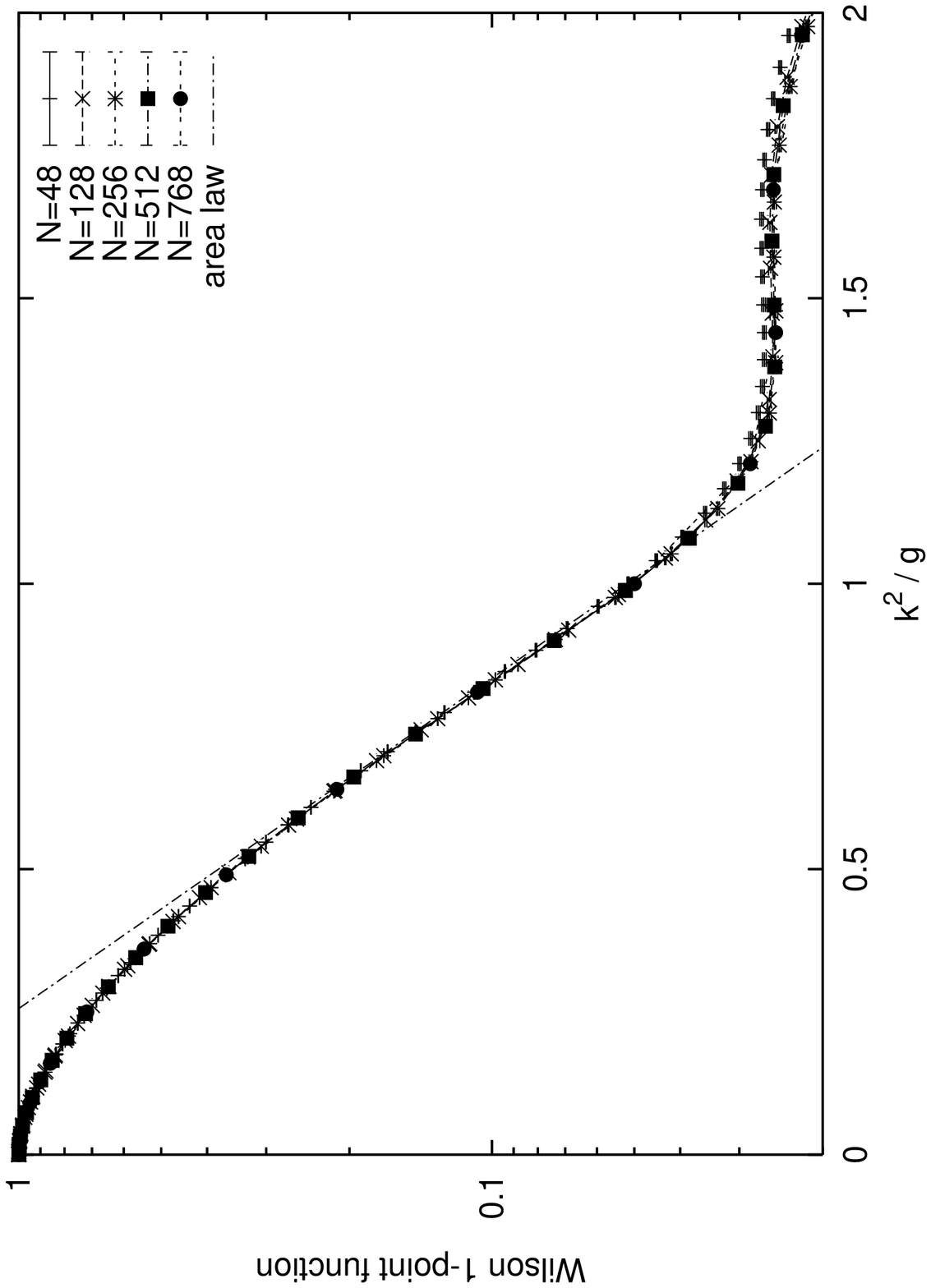}
\caption{\it{The Wilson loop plotted logarithmically against
$k_{\rm phys}^{2}=k^{2} / g$. 
The regime $k^{2}_{\rm phys} \approx 0.5 \dots 1$ 
is consistent with an area law.}}
\label{arealaw-fig}
\end{figure}

In Fig.\ \ref{arealaw-fig} we show the Wilson loop logarithmically 
against $k^{2}_{\rm phys}$, and we see a window, ranging approximately 
from $k^{2}_{\rm phys} \approx 0.5 \dots 1$, which is in
agreement with the area law. 
This area law regime turns out to be manifestly stable
as $N$ becomes large. Therefore, we do observe a finite range of
scale where Eguchi-Kawai equivalence {\em is} in business.

\begin{figure}[hbt]
\def\fpsangle{270}
\epsfxsize=100mm
\fpsbox{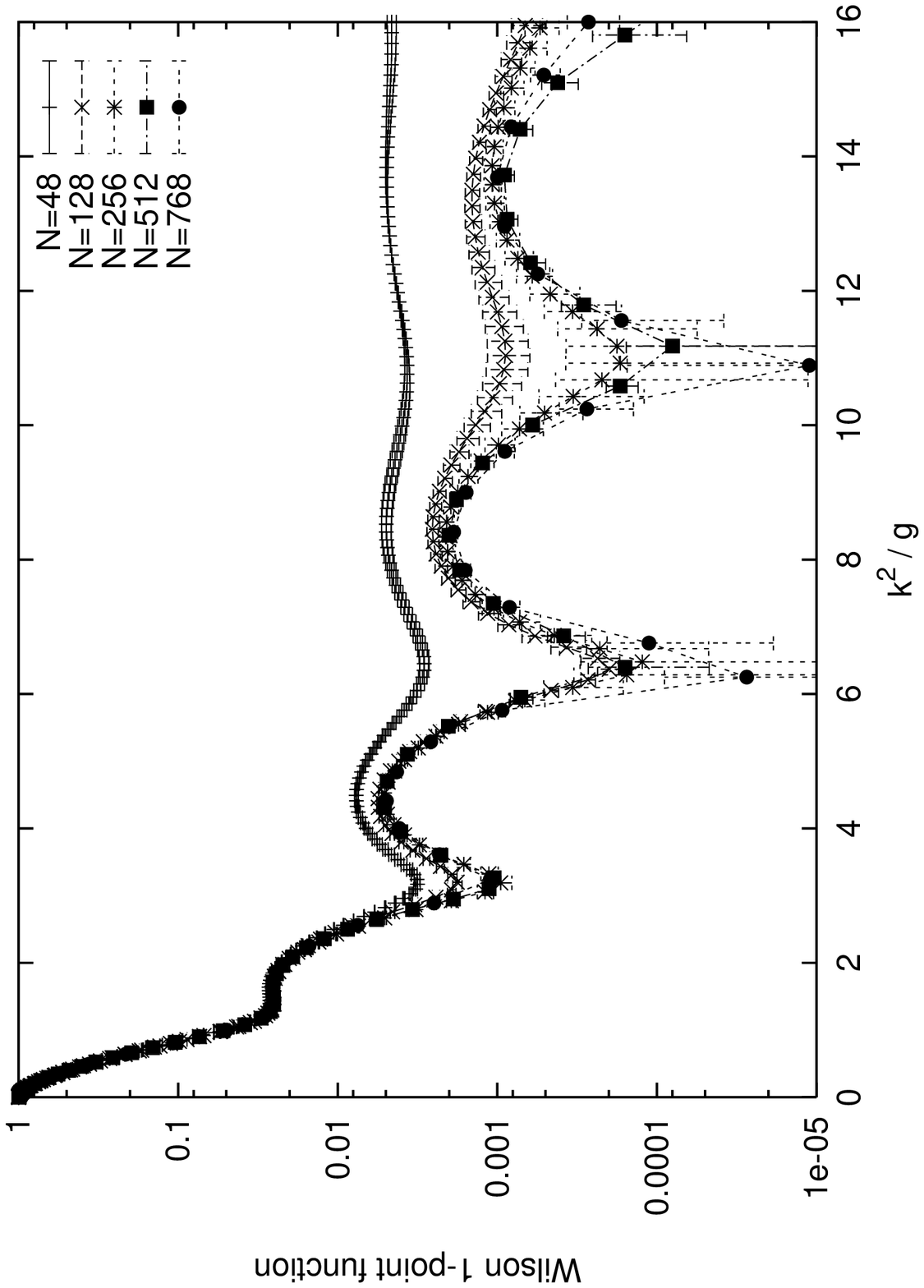}
\caption{\it{The Wilson loop plotted against
$k_{\rm phys}^{2}=k^{2} / g$.
There seems to be a sequence of linearity windows at large
$N$, but the relevant area law window is the first one,
$k^{2}/g \approx 0.5 \dots 1$.}}
\label{wil-log-fig}
\end{figure}

In Fig.\ \ref{wil-log-fig} we extend this plot up to quite
large momenta in order to check if perhaps a further area law
regime shows up. We observe an oscillating behavior, which
also persists at large $N$. These oscillations might include
additional, small area law regimes, but the relevant 
interval with this respect is clearly the first one, shown in Fig.\ 
\ref{arealaw-fig}.
We also looked for a regime where the perimeter
law holds, but we observed that there is no such interval 
with a size comparable to the leading area law regime.

\section{Conclusions}

In view of the literature on the Eguchi-Kawai model, 
it comes as a surprise that the area law does hold
in a significant range of scale for the bosonic IIB matrix
model. It implies that Eguchi-Kawai equivalence
may hold in this regime, even in the absence of twists
or other mechanisms to preserve the phase symmetry.

This observation opens up interesting perspectives, 
with respect to both, the theoretical background and
applications. It is now motivated to reconsider the
possibility of Eguchi-Kawai equivalence also in the original
Eguchi-Kawai model, or further variants thereof,
which do not involve a symmetry preserving mechanism.
In the matrix model studied here one could still
investigate whether the area law regime extends to infinity
if we modify the model by the quenching procedure \cite{SCRI,GrKi}.
This question is non-trivial given that the Hermitian matrix model has no
direct link to lattice gauge theory, and its Eguchi-Kawai equivalence
has been discussed only at the perturbative level \cite{GrKi}.
Such a study may also shed light on the string-theoretical meaning 
of the quenching procedure.

We hope that our results presented here provide new insight
towards a string description of large $N$ QCD.

\section*{Acknowledgment}
The authors would like to thank J.\ Ambj\o rn, H. Kawai and P. Olesen
for stimulating discussions.  K.N.A.'s research was partially
supported by RTN grants HPRN-CT-2000-00122, HPRN-CT-2000-00131 and
HPRN-CT-1999-00161 and the INTAS contract N 99 0590. W.B.
was partially supported by the EU-RTN network under contract
FMRX-CT97-0122.

\end{document}